\documentstyle[prd,aps,epsf]{revtex}

\begin{document}

\twocolumn[\hsize\textwidth\columnwidth\hsize\csname
@twocolumnfalse\endcsname

\title{Stability of ${\rm AdS}_p 
\times {\rm S}^n \times {\rm S}^{q-n}$ Compactifications}

\author{Tetsuya Shiromizu$^{(a)}$, 
Daisuke Ida$^{(a)}$, Hirotaka Ochiai$^{(b)}$ 
and Takashi Torii$^{(a), (c)}$}

\address{$^{(a)}$ Research Centre for the Early Universe (RESCEU),
The University of Tokyo, Tokyo 113-0033, Japan}
\address{$^{(b)}$
Department of Physics, The University of Tokyo, Tokyo 113-0033, Japan}
\address{$^{(c)}$ Advanced Research Institute for Science and Engineering,
Waseda University,
Shinjuku-ku, Tokyo 169-8555, Japan}



\maketitle

\begin{abstract}
We examine the stability of ${\rm AdS}_p \times {\rm S}^n 
\times {\rm S}^{q-n}$.
The initial data constructed by De Wolfe et al \cite{Gary} has been
carefully analyised and we have confirmed that there is no lower bound for 
the total mass for $q< 9$. 
The effective action on ${\rm AdS}_p$ has been
derived for dilatonic compactification of the system
to describe the non-linear fluctuation of the background space-time.
The stability is discussed applying
the positive energy theorem to the effective theory on AdS, which again
shows the stability for $q \geq 9$.
\end{abstract}
\vskip2pc]

\vskip1cm

\section{Introduction}\label{I}

Related to the bosonic M-theory\cite{Bosonic}, 
there is a much interest in the stability of 
${\rm AdS}_p \times {\rm M}_q$ space-times with $p+q$
dimensions higher than eleven. This type of 
space-time appears through so-called 
Freund-Rubin compactifications \cite{FR} or 
near horizon geometry of extreme black-brane solutions.  
{}From the AdS/CFT correspondence\cite{AdSCFT} including holographic 
renormalisation\cite{RG} or 
the brane-world point of view\cite{Lisa}, the stability of this system is 
a fundamental and important issue. In addition, the understanding 
of de Sitter (dS) space-time in string theory might be able to be 
accomplished by using AdS/CFT correspondence \cite{deSitter,Stro}:
How can we explain the dS entropy in the string theory?  
Therein our universe is assumed to be ${\rm dS}_q$ after Wick 
rotation of ${\rm S}^q$ of ${\rm AdS}_p \times {\rm S}^q$. 
In the sense of Ref \cite{deSitter}, this means that 
the stability of our space-time might be related to that of 
the full space-time, ${\rm AdS}_p \times {\rm S}^q$, which is known to 
be stable. 

It is well known 
that the full space-time will in general be unstable for 
${\rm M}_q$ having non-trivial topology (For example, see 
old references \cite{Yasuda}). This type of instability frequently 
happens in asymptotically flat space-times, for example, 
Kaluza-Klein bubble is famous one \cite{Witten,Brill}. 
The recent analysis 
of ${\rm AdS}_p  \times 
{\rm S}^n \times {\rm S}^{q-n}$ shows, however, 
stability against linear perturbations for $q \geq 9$ \cite{Gary}. 
They have also performed a non-linear realisation
of stability in terms of the initial data  \cite{Gary}.
Furthermore,
it is shown that the energy become arbitrary negative regardless of 
the size of the compactified space for unstable cases ($q<9$).
These discussions indicate that the instability of the higher 
dimensional universe, 
such as ${\rm dS}_4 \times {\rm S}^{q-4}$, depends on the dimension $q$ 
in the sense of Ref. \cite{deSitter}.   
Here ${\rm dS}_4 \times {\rm S}^{q-4}$ is assumed to realise as 
the Wick rotated ${\rm AdS}_p \times {\rm S}^{4} \times {\rm S}^{q-4}$ 
space-time. 

In this paper, first of all, we reanalyse in detail the feature of the 
initial data presented in Ref \cite{Gary} in Sec.~\ref{II}.
We evaluate the critical dimension of the stability in several cases. 
Due to non-linear effects, the critical dimension might be different 
from that expected by the linear 
perturbative analysis. 
In Sec.~\ref{III}, we consider the dilatonic compactifications of 
${\rm AdS}_p \times {\rm S}^{n} \times {\rm S}^{q-n}$
and derive the effective action on ${\rm AdS}_p$ to describe the 
non-linear effect of fluctuations, which can also describe the
subsequent evolution of the initial data. 
Due to the curvature of 
the compactified space, ${\rm S}^n \times {\rm S}^{q-n}$, we have 
non-trivial effective potential for the dilatonic field. In principle, we 
can use the potential to justify the stability. The potential, however, 
does not have the lower bound. Then, we might be able to conclude 
that the system is always unstable, though this is not the case, since
there are stable tachyonic fields in AdS \cite{BZ}. 
In Sec.~\ref{IV}, we discuss this issue at the non-linear level 
by using the positive energy 
theorem for asymptotically ${\rm AdS}_p$ space-times. 
Our result shows that the system will be stable for $q\ge 9$.
Finally, we give the summary in Sec.~\ref{V}. 

\section{The initial data}\label{II}

\subsection{Freund-Rubin compactifications}

We begin with the action
%
\begin{eqnarray}
S_{p+q}=\int d^{p+q}x \sqrt{|g|}\left(\frac{1}{2} R
-\frac{1}{4q!}
{F_q}^2 \right),
\label{starting}
\end{eqnarray}
%
where $R$ 
is the ($p+q$)-dimensional Ricci scalar and 
$F_q$ is the field strength of the 
$q$-form field. The field equations are
given by
%
\begin{equation}
dF_q=0,~~~d*F_q=0,
\label{eq-form}
\end{equation}
and
\begin{eqnarray}
R_{MN} & = & \frac{1}{2(q-1)!}F_{MI_2 \cdots I_q}F_N^{~~I_2 \cdots I_q} 
\nonumber \\
& & ~~~-\frac{q-1}{2(p+q-2)q!} g_{MN}{F_q}^2.
\end{eqnarray}
%
This system admits the Freund-Rubin (FR) solution \cite{FR}.
The metric of the FR solution consists of the direct
sum of the $p$- and $q$-dimensional Einstein space metrics:
%
\begin{eqnarray}
ds^2=g_{\mu\nu}dx^\mu dx^\nu+g_{ij}dx^i dx^j,
\end{eqnarray}
%
where $g_{\mu\nu}$ and $g_{ij}$ are the metrics of the $p$- and 
$q$-dimensional Einstein spaces satisfying 
%
\begin{eqnarray}
{}^{(p)}\!R_{\mu\nu}=-\frac{(p-1)}{L^2}g_{\mu\nu}
\end{eqnarray}
%
and
%
\begin{eqnarray}
{}^{(q)}\!R_{ij}=\frac{(q-1)}{R^2}g_{ij},
\end{eqnarray}
%
respectively.
The field strength is given by
%
\begin{eqnarray}
F_q=f *_q1,
\end{eqnarray}
%
where $f$ is a constant which is related to the curvature 
radii $L$ and $R$ as 
%
\begin{eqnarray}
f^2=\frac{2(p+q-2)(q-1)}{(p-1)R^2}=\frac{2(p+q-2)(p-1)}{(q-1)L^2}, \label{cr}
\end{eqnarray}
%
and $*_q1$ denotes the volume form of the $q$-dimensional Einstein space.

According to the linear perturbation analysis around FR solutions 
\cite{Gary}, 
${\rm AdS}_p \times {\rm S}^q$ is always 
stable. On the other hand, the situation 
changes when the $q$-dimensional space has non-trivial topology;
For ${\rm AdS}_p \times 
{\rm S}^n \times {\rm S}^{q-n}$ with $q \geq 9$, the system is stable, while 
unstable for $q<9$. We shall call $q=9$ the critical dimension. 
It is also considered an initial data of ${\rm AdS}_p \times 
{\rm S}^n \times {\rm S}^{q-n}$ 
 to demonstrate 
the instability of the system at non-linear level. 
We will describe and carefully anaylse the initial data
in the next subsection. 

\subsection{Time-symmetric initial data}

The time-symmetric initial data considered in Ref. \cite{Gary}
is assumed to have the following form
%
\begin{eqnarray}
d\ell^2 & = & \frac{1}{h(r)}dr^2+r^2d \Omega_{p-2}^2 \nonumber \\
& & +e^{(q-n)\phi (r)}d\Omega_{n}^2+e^{-n \phi (r)}d\Omega_{q-n}^2,
\end{eqnarray}
%
where 
%
\begin{eqnarray}
h(r)=1-\frac{m(r)}{r^{p-3}}+\frac{r^2}{L^2},
\end{eqnarray}
%
and $d\Omega_n^2$ and $d\Omega_{q-n}^2$ are the metrics of 
$S^n$ and $S^{q-n}$,
respectively. 
If $\phi$ approaches zero sufficiently fast 
as $r \to +\infty$, the hypersurface is expected to be 
asymptotically ${\rm AdS}_p \times {\rm S}^n \times {\rm S}^{q-n}$. 

Since we are considering the time-symmetric initial data,
the equation to be solved is just the Hamiltonian constraint
%
\begin{eqnarray}
(p-2)\frac{m'(r)}{r^{p-2}} & = & \frac{1}{4}qn(q-n)h(r) \phi'^2 
-\frac{q-1}{R^2}\left[ ne^{-(q-n)\phi (r)} \right.\nonumber \\
& &\left. ~~~+(q-n)e^{n\phi (r)}-q\right].\label{m'}
\end{eqnarray}
%
The momentum constraint is automatically satisfied. 

\begin{figure}[t]
\vspace*{1mm}
\begin{center}
\epsfxsize=2.7in
~\epsffile{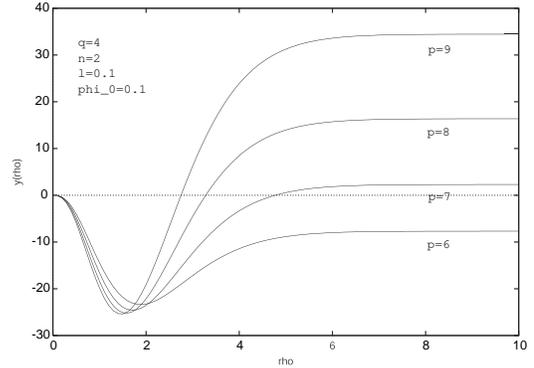}
\end{center}
\caption{The mass function $y(\rho)$ with parameters
  $p=6,7,8,9$, $q=4$, $n=2$, $\ell=0.1$, $\phi_0=0.1$.}
\label{fig1}
\end{figure}

We set $\phi (r)=\phi_0 e^{-r/a}$ $(\phi_0={\rm constant})$ and 
numerically solve the equation for $m(r)$
with the regular boundary condition 
at the center, 
$m(r \sim 0)=O(r^{\lambda})$, where $\lambda\geq p-1$. 
This regularity condition is important 
when one argues the positivity of the total mass 
(c.f.  Schwarzschild solution with $M<0$).
For convenience' sake, we introduce dimensionless variables:
\begin{eqnarray}
\ell:=L/a,~~
\rho:=r/a,~~
y:=m/L^{p-3}.
\end{eqnarray}
Then, Eq.~(\ref{m'}) becomes
\begin{eqnarray}
&&(p-2)\frac{\ell^{p-3}}{\rho^{p-2}}y'(\rho)  =  \frac{qn(q-n)}{4}
\left[1-\frac{\ell^{p-3}}{\rho^{p-3}}y(\rho)+\frac{\rho^2}{\ell^2}\right]
\phi^2\nonumber\\
&&-\frac{(p-1)^2}{\ell^2(q-1)}\left[ ne^{-(q-n)\phi}
+(q-n)e^{n\phi }-q\right].\label{y'}
\end{eqnarray}
This system is characterized by five parameters $\{p,q,n,\ell,\phi_0\}$.
Typical behavior of the mass function $y(\rho)$ is shown in Fig.~\ref{fig1}.
We have computed the asymptotic values of the mass function 
$y_\infty:=\lim_{\rho\rightarrow+\infty}y(\rho)$ for a sufficiently wide 
range of parameters, and confirmed that a negative mass solution can arise 
only when $q<9$.
The tables~\ref{critical1}-\ref{critical4}
shows this critical dimension $q=9$.
We can see from these tables that a negative mass solution likely arises when
(i) $p$ is small, 
(ii) $q$ is small,
and
(iii) $n$ is close to $q/2$.
For $\ell\ll 1$, Eq.~(\ref{y'}) takes the form
\begin{equation}
y'\simeq\varphi(p,q,n,\phi_0,\rho)\ell^{-p+1},
\end{equation}
which implies that $y_\infty\propto \ell^{-p+1}$ for fixed values of the 
other parameters.
Hence if we find a negative mass solution for $\ell\ll 1$, then we can also 
obtain 
solutions with arbitrarily large negative mass, which indicates the
instability of the system.

Here we must have a minor comment on the evaluation of the 
critical dimension. For $\phi \ll 1$, Eq.~(\ref{m'}) becomes
%
\begin{eqnarray}
\frac{p-2}{qn(q-n)} \frac{m'(r)}{r^{p-2}} \simeq h(r)\phi'^2
-2\frac{(p-1)^2}{L^2(q-1)}\phi^2.\label{hami2}
\end{eqnarray}
%
In Ref. \cite{Gary} there is a rough argument to evaluate 
the critical dimension by assuming the positive signature of 
$m(r)$ in $h(r)$ in the right-hand side. However, $m(r)$ 
does not have a definite signature.

\section{Effective theory on AdS}\label{III}

{}From now on we derive the effective theory which describes the time 
evolution of the initial data presented in the previous section. 
To do so, we consider only zero-modes. Since we are interested in the 
dynamics of the full space-time, we adopt the following standard 
dimensional reduction:
%
\begin{equation}
ds^2 =g_{\mu\nu}(x)dx^\mu dx^\nu
+e^{2\phi_1(x)} d\Omega_n^2+e^{2\phi_2(x)} d\Omega_{q-n}^2.
\label{gfr-metric}
\end{equation}
%
The curvature radii of $S^n$ and $S^{q-n}$ are normalized such 
that
Eq.~(\ref{gfr-metric}) gives the Freund-Rubin metric when $\phi_1=\phi_2=0$.
Then, the following form of the $q$-form field
%
\begin{equation}
F_q=f *_n1\wedge *_{q-n}1,
\end{equation}
%
where $*_n1$ and $*_{q-n}1$ are the volume forms of 
$S^n$ and $S^{q-n}$, respectively, solves the field equations
(\ref{eq-form}).
%
%
Then, the Einstein equation becomes
%
\begin{eqnarray}
{}^{(p)}\!R_{\mu\nu}&=&-\frac{(p-1)}{L^2}e^{-2\psi}
g_{\mu\nu}+\nabla_\mu\nabla_\nu\psi\nonumber\\
&&+n\nabla_\mu\phi_1\nabla_\nu\phi_1+(q-n)\nabla_\mu\phi_2\nabla_\nu\phi_2,
\label{gfr-einstein}
\end{eqnarray}
%
and
%
\begin{equation}
\nabla^2\Phi^X+\nabla_\mu\psi\nabla^\mu\Phi^X
=\frac{(q-1)}{R^2}(e^{-2\Phi^X}-e^{-2\psi}),\label{gfr-phi}
\end{equation}
%
where $\nabla$ denotes the covariant derivative with respect to 
$g_{\mu\nu}$, $\Phi^X$ represents $\phi_1$ and $\phi_2$, and
%
\begin{equation}
\psi:=n\phi_1+(q-n)\phi_2
\end{equation}
%
has been defined. 
To eliminate terms including second derivative of $\psi$ in 
Eq.~(\ref{gfr-einstein}),
we perform the conformal transformation:
%
\begin{equation}
g_{\mu\nu}\mapsto \bar{g}_{\mu\nu}=e^{2\psi/(p-2)}g_{\mu\nu}
\end{equation}
%
Then, Eqs.~(\ref{gfr-einstein}), (\ref{gfr-phi}) become
%
\begin{eqnarray}
{}^{(p)}\!\bar{R}_{\mu\nu}&=&\frac{n(p+n-2)}{p-2}
\bar{\nabla}_\mu\phi_1\bar{\nabla}_\nu\phi_1
\nonumber \\
&+&\frac{2n(q-n)}{p-2}\bar{\nabla}_{(\mu}\phi_1\bar{\nabla}_{\nu)}\phi_2
\nonumber \\
&+&\frac{(q-n)(p+q-n-2)}{p-2}\bar{\nabla}_\mu\phi_2\bar{\nabla}_\nu\phi_2
\nonumber\\
&+&\frac{f^2}{2(p-2)}
\biggl\{e^{-\frac{2(p-1)}{p-2}\psi}
\nonumber\\
&&-\frac{(p-1)[ne^{-2\phi_1}+(q-n)e^{-2\phi_2}]}{(p+q-2)}
e^{-\frac{2}{p-2}\psi}\biggr\}\bar{g}_{\mu\nu},
\nonumber\\
\end{eqnarray}
\begin{equation}
\bar{\nabla}^2\Phi^X+\frac{f^2(p-1)}{2(p+q-2)}\left\{
e^{-\frac{2(p-1)}{p-2}\psi}\right.
\left.-e^{-2\Phi^X}
e^{-\frac{2}{p-2}\psi}
\right\}=0.
\end{equation}
%
These are derived by variation of the effective action
in Einstein frame:
%
\begin{equation}
\bar{S}=\int d^px\sqrt{|\bar{g}|}\left[
\frac{1}{2}{}^{(p)}\!\bar{R}-\frac{1}{2}G_{XY}
\bar{g}^{\mu\nu}{\Phi^X}_{\!\!,\mu}{\Phi^Y}_{\!\!,\nu}
-V(\Phi)\right],
\end{equation}
%
where
%
\begin{eqnarray}
G&=&\frac{n(p+n-2)}{p-2}d{\phi_1}^{\!\!2}
+\frac{2n(q-n)}{p-2}d\phi_1 d\phi_2\nonumber\\
&&+\frac{(q-n)(p+q-n-2)}{p-2}d{\phi_2}^{\!\!2}
\end{eqnarray}
%
and
%
\begin{eqnarray}
V(\Phi)&=&\frac{f^2}{4}
e^{-\frac{2(p-1)}{p-2}\psi}
-\frac{f^2(p-1)}{4(p+q-2)}\nonumber\\
&&\times[ne^{-2\phi_1}+(q-n)e^{-2\phi_2}]
e^{-\frac{2}{p-2}\psi}.
\label{epot}
\end{eqnarray}
%

Unstable modes in Ref. \cite{Gary} corresponds to 
the $\psi =0$ case with the potential
%
\begin{eqnarray}
V(\Phi)&=&\frac{f^2}{4}\left\{1
-\frac{p-1}{4(p+q-2)}\right.
\nonumber\\
&&\times\left[ne^{-2\phi_1}+(q-n)e^{\frac{2(q-n)}{n}\phi_1}
\right]\biggr\}
\end{eqnarray}
%

\begin{figure}[t]
\vspace*{-2mm}
\begin{center}
\epsfxsize=2.7in
~\epsffile{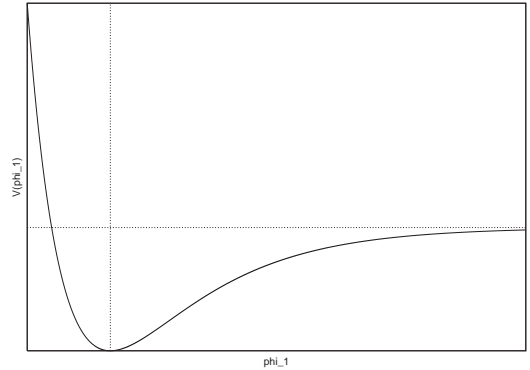}
\end{center}
\caption{This is the typical profile of the potential $V(\Phi)$ for 
${\rm AdS}_p \times {\rm S}^q$. It is definitely stable.}
\label{fig3}
\end{figure}

For ${\rm AdS}_p \times {\rm S}^q$ the potential can be 
obtained by setting $n=q$ and $\phi_2=0$ ($\psi = q \phi_1$);
%
\begin{eqnarray}
V(\Phi)=\frac{f^2}{4}\left[e^{-\frac{2q(p-1)}{p-2}\phi_1}
-\frac{q(p-1)}{p+q-2}
e^{-\frac{2(p+q-2)}{p-2}\phi_1} \right].
\end{eqnarray}
%
This has the lower bound $V(0)=-(p-1)(p-2)/2L^2$ at $\phi_1=0$
and the  system is definitely stable (See Fig.~\ref{fig3}). 

At first glance, the system seems to be unstable because the 
potential (\ref{epot}) does not have the lower bound. However, we must 
carefully analyse the stability because we are thinking of the 
asymptotically AdS space-times, in which case a tachyonic
field may be stable \cite{BZ}. Correspondingly, the system can be 
stable if the potential satisfies a certain condition \cite{Town}. 

\section{Implications from the positive energy theorem}\label{IV}

In this section, we discuss the stability of the ${\rm AdS}_p 
\times {\rm S}^{n} \times {\rm S}^{q-n}$ space-time by applying
the positive energy theorem \cite{Town,PET} to the 
effective theory on ${\rm AdS}_p$.

If the scalar potential can be written in the form
%
\begin{equation}
V = 2(p-2)^2G^{XY}\frac{\delta W}{\delta \Phi^X}\frac{\delta W}{\delta \Phi^Y}
 -2(p-1)(p-2)W^2, \label{super}
\end{equation}
%
we can show the positivity of the total energy in the Einstein 
frame \cite{Town}. Indeed, the total energy\footnote{This energy 
is identical to that defined by Abbott and Deser in 
asymptotically AdS space-times\cite{AD,AdS}. For the 
trivial potential, $V={\rm const.}$, or supergravity theory, 
see \cite{GH2P,Gibbons}.} can be expressed as
%
\begin{eqnarray}
M_{\rm AD}=\int_\Sigma d^{p-1}x 
\biggl[2 |\hat D \epsilon|^2 + \sum_X |\delta
 \lambda^{(X)}|^2
\biggr], 
\label{energy}
\end{eqnarray}
%
where $\Sigma$ is the asymptotically ${\rm AdS}_p$ space-like hypersurface,
$\hat D_\mu :=D_\mu +iW (\Phi) \gamma_\mu$ with the spinor covariant 
derivative $D_\mu$, and $\epsilon$ is the spinor satisfying the 
Witten equation\footnote{The existence of this solution 
should be able to be confirmed in the similar way as 
Ref. \cite{Gibbons}}, $\sum_{a=1}^{p-1}\gamma^a \hat D_a \epsilon =0$. 
The spinor $\delta \lambda^{(X)}$ is defined by
%
\begin{equation}
\delta \lambda^{(X)}=\frac1{\sqrt{2}}
\left[i f^{(X)}_Y \gamma^\mu \partial_\mu 
\Phi^Y
+2(p-2)f^{(X)Y}\frac{\delta W}{\delta \Phi^Y} \right]\epsilon,
\end{equation}
%
where $f^{(X)}_Y$ is such that
%
\begin{equation}
G_{XY}=\sum_Zf^{(Z)}_Xf^{(Z)}_Y
\end{equation}
%
holds, which can be thought of as the orthonormal basis of the target space.

For simplicity, we consider the situation in which 
the effective theory is described by a single scalar field $\Phi$
defined by
%
\begin{equation}
\phi_1=\frac{\alpha}{n}\Phi,~~~\phi_2=\frac{\beta}{q-n}\Phi.
\end{equation}
%
By the scaling property of the scalar field, we can put
%
\begin{equation}
\alpha +i\beta=e^{i\theta},~~~(0\leq\theta<\pi),
\end{equation}
%
without loss of generality.
This contains both of the typical stable and unstable systems. 
The parameter adopted in Sec.~\ref{II}B corresponds to
$\alpha=-\beta$, 
$(\theta=3\pi/4)$.
The target space metric and the effective potential become
%
\begin{eqnarray}
G&=&\left[\frac{(\alpha+\beta)^2}{p-2}+\frac{\alpha^2}{n}
+\frac{\beta^2}{q-n}\right]d\Phi^2
\end{eqnarray}
%
and
%
\begin{eqnarray}
V(\Phi)&=&\frac{f^2}{4}e^{-\frac{2(p-1)(\alpha+\beta)}{p-2}\Phi}
\nonumber\\
&&-\frac{f^2(p-1)}{4(p+q-2)}[ne^{-2\alpha\Phi/n}+(q-n)e^{-2\beta\Phi/(q-n)}]
\nonumber\\
&&\times e^{-\frac{2(\alpha+\beta)}{p-2}\Phi}.
\end{eqnarray}
%
The asymptotically AdS condition requires that the scalar field
 approaches  $\Phi=0$ at spatial infinity,
where the effective potential has a stationary point. To 
confirm whether there exist a functional $W(\Phi)$
satisfying Eq.~(\ref{super}), we expand the effective potential in 
terms of $\Phi$ under the assumption $|\Phi| \ll 1$.
Up to the second order in $\Phi$, the effective potential becomes
%
\begin{equation}
V(\Phi)=V_0+V_1\Phi+\frac{1}{2}V_2\Phi^2+O(|\Phi|^3),
\end{equation}
%
where
%
\begin{eqnarray}
V_0&=&-\frac{f^2(p-2)(q-1)}{4(p+q-2)},\\
V_1&=&0,\\
V_2&=&\frac{f^2(p-1)}{(p+q-2)}\nonumber\\
&&\times\left[\frac{(\alpha+\beta)^2(p+q-3)}{p-2}
-\frac{\alpha^2}{n}-\frac{\beta^2}{q-n}\right].
\end{eqnarray}
%
The functional $W(\Phi)$ is also expanded as
%
\begin{equation}
W(\Phi)=W_0+W_1\Phi+\frac{1}{2}W_2\Phi^2+O(|\Phi|^3). \label{exp}
\end{equation}
%
{}From Eq.~(\ref{super}) we find
%
\begin{eqnarray}
W_0&=&\pm \frac{f}{2}\sqrt{\frac{q-1}{2(p-1)(p+q-2)}},\\
W_1&=&0,
\end{eqnarray}
%
and $W_2$ is determined by solving the following algebraic
equation:
%
\begin{equation}
G^{\Phi \Phi}{W_2}^2-\frac{W_0(p-1)}{p-2}W_2
-\frac{1}{4(p-2)^2}V_2=0,
\label{quadraW}
\end{equation}
%
which has real solutions if
%
\begin{equation}
\frac{(\alpha+\beta)^2(8p+9q-25)}{p-2}
+\left(\frac{\alpha^2}{n}+\frac{\beta^2}{q-n}\right)(q-9)\ge0
\end{equation}
%
is satisfied. The condition that the above inequality holds
for every values of $\alpha$, $\beta$ is
%
\begin{equation}
\frac{(p+q-2)(q-9)(q-1)}{n(q-n)(p-2)}\ge0.
\end{equation}
%
This means that the total energy is always positive
for small fluctuation of $\Phi$ if $q\ge 9$ is satisfied, 
which is consistent with the linear analysis in Ref.~\cite{Gary}.
When $q=9$, a double real root appears for 
$\theta=3\pi/4$, of which direction
the potential $V$ shows negative mass squared.

\begin{figure}[t]
\vspace*{-5mm}
\begin{center}
\epsfxsize=3.2in
~\epsffile{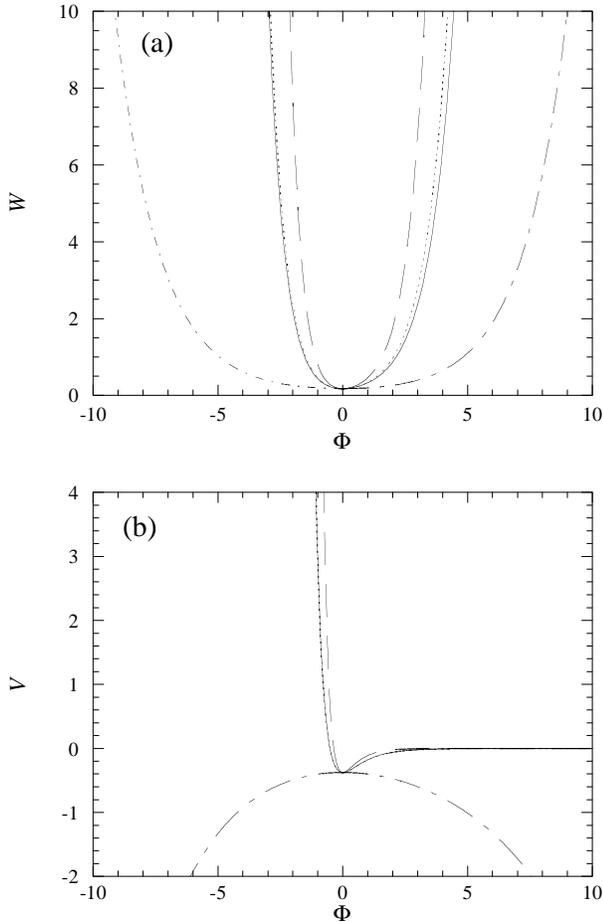}
\end{center}
\caption{(a) The functional $W$ and (b) the potential $V$
of the scalar field with the parameter 
$(p, q, n)=(4, 10, 4)$ and $\theta=0$ (solid line), $\pi/4$
(dashed line), $\pi/2$ (dotted line), $3\pi/4$ 
(dot-dashed line).
Even if the potential shows negative mass squared when 
$\theta=3\pi/4$, we can find $W$.}
\label{fig4}
\end{figure}

For large fluctuations of $\Phi$, we need numerical calculations
to confirm the existence of the functional $W(\Phi)$. This seems to be 
almost promising work, because we can have the asymptotic solutions 
of $W(\Phi)$ at $\Phi \to \pm \infty$ and all the higher order terms 
for $|\Phi| \ll 1$ as Eq. (\ref{exp}). 
The numerical analysis shows that there is always the 
functional $W(\Phi)$ when $q\geq 9$. Some cases are shown in 
Fig.~\ref{fig4}. 

Thus, we expect that the ${\rm AdS}_p\times{\rm S}^n\times{\rm S}^{q-n}$ 
for $q\ge 9$ is
stable even for large fluctuations.
The advantages of the use of the positive energy theorem are 
(i) one can  confirm stability of the background geometry 
beyond linear perturbative analysis, 
(ii) one can argue the semi-classical or quantum mechanical  stability without
topology change of the space-time,
and
(iii) the background geometry need not be exactly
${\rm AdS}_p\times{\rm S}^n\times{\rm S}^{q-n}$; the positive energy theorem
can be applied to the case of the non-trivial spatial topology such as
black hole space-times \cite{GH2P}.


\section{Summary}\label{V}

In this paper we examined the stability of ${\rm AdS}_p \times {\rm S}^n 
\times {\rm S}^{q-n}$, that is, generalised Freund-Rubin compactifications. 
We confirmed that there are initial data having 
the arbitrary negative total mass for $q < 9$. 

We have constructed the effective theory on AdS
and then considered the condition  
that the positivity of the total mass can be confirmed. 
We can expect that the ${\rm AdS}_p
\times{\rm S}^n\times{\rm S}^{q-n}$ for $
q\ge 9$ is
also stable under quantum mechanical or non-linear fluctuations.
Furthermore, this argument will also hold for AdS-black hole background, 
which might be relevant for the stability of  some thermal 
${\rm CFT}_{p-1}$ according
to the AdS/CFT correspondence.

We are interested in the time evolution of the initial data with the 
negative energy. In the case of Kaluza-Klein bubble with the negative 
energy, it is remembered that 
we can see the tendency of the appearance of the naked 
singularity, although we cannot have the definite answer since this relies on
numerical analysis \cite{Shinkai}. 
Similarly, such a problem for the present situation will be also shared.
We need systematic and numerical analysis to clarify this point.

\section*{Acknowledgments}

We would like to thank Y. Shimizu and K. Takahashi for fruitful discussions. 
TS's work is partially supported by Yamada Science Foundation.


\newpage

\begin{table}[t]
\begin{center}
\begin{tabular}{c|c c c }
\noalign{\hrule height0.8pt}
$q \backslash n$ & $2$ & $3$ & $4$ \\
\hline
4 & 4 & * & * \\
5 & 5 & * & * \\
6 & 5 & 5 & * \\
7 & 7 & 6 & * \\
8 & 11 & 10 & 9 \\
9 & + & + & + \\
\hline
\end{tabular}
\end{center}
\caption{$\ell=0.01$, $\phi_0=1$ case. The figures displayed denote  
the maximum dimension $p$ for the negative mass. $+$ means that the 
mass is positive. $*$ denotes redundant cases.}
\label{critical1}
\end{table}

\begin{table}[t]
\begin{center}
\begin{tabular}{c|c c c}
\noalign{\hrule height0.8pt}
$q \backslash n$ & $2$ & $3$ & $4$\\
\hline
4 & 4 &* & *  \\
5 & 4 & * & * \\
6 & 5 & 5 & * \\
7 & 7 & 6 & * \\
8 & 11 & 10 & 9 \\
9 & + & + & + \\
\hline
\end{tabular}
\end{center}
\caption{$\ell=0.1$, $\phi_0=1$.
}
\label{critical2}
\end{table}

\begin{table}[t]
\begin{center}
\begin{tabular}{c|c c c }
\noalign{\hrule height0.8pt}
$q \backslash n$ & $2$ & $3$ & $4$ \\
\hline
4 & 4 &* & * \\
5 & 5 & * & * \\
6 & 6 & 6 & * \\
7 & 8 & 8 & * \\
8 & 12 & 12 & 11 \\
9 & + & + & + \\
\hline
\end{tabular}
\end{center}
\caption{$\ell=1$, $\phi_0=1$.
}
\label{critical3}
\end{table}

\begin{table}[t]
\begin{center}
\begin{tabular}{c|c c c}
\noalign{\hrule height0.8pt}
$q \backslash n$ & $2$ & $3$ & $4$ \\
\hline
4 & 4 &* & * \\
5 & 4 & * & * \\
6 & 4 & 4 & * \\
7 & 5 & 5 & * \\
8 & 9 & 9 & 9 \\
9 & + & + & + \\
\hline
\end{tabular}
\end{center}
\caption{$\ell=0.01$, $\phi_0=0.01$.
}
\label{critical4}
\end{table}

\end{document}